\documentclass[aps,prb,twocolumn,longbibliography]{revtex4-2}

\usepackage[utf8]{inputenc}
\usepackage{graphicx}
\usepackage{hyperref}
\usepackage{xcolor,comment}
\usepackage{natbib}
\usepackage[T1]{fontenc}

\usepackage{amsmath,amsthm,amssymb}
\usepackage{mathtools}
\usepackage{braket}
\usepackage{bbm,bm}
\usepackage[bb=boondox,scr=boondox]{mathalfa}

\newcommand{\ie}{{i.e.},\ }

\newcommand{\tio}{\mathscr{O}}

\begin{document}

\title{Eigenstate~thermalization~for~local~versus~translationally~invariant~observables}

\author{Rohit Patil}
\affiliation{Department of Physics, The Pennsylvania State University, University Park, Pennsylvania 16802, USA}
\author{Marcos Rigol}
\affiliation{Department of Physics, The Pennsylvania State University, University Park, Pennsylvania 16802, USA}

\begin{abstract}
Local observables and their translationally invariant counterparts are generally thought to provide the same predictions for experiments. While this equivalence holds for expectation values in clean systems (up to finite-size effects), it is often assumed to extend to correlation functions, where it need not hold. We examine this assumption from the perspective of the eigenstate thermalization hypothesis. Specifically, we explore the spectral functions of local and translationally invariant observables in the spin-1 tilted-field Ising chain with periodic and open boundary conditions. We identify the contexts in which these observables and boundary conditions differ and those in which they are interchangeable. We unveil an off-diagonal eigenstate thermalization in translationally invariant systems for matrix elements between energy eigenstates with different quasimomenta.
\end{abstract}

\maketitle

\emph{Introduction.}
Experiments often probe local observables in the bulk of a material, for example, charge and spin correlations in a high-temperature superconductor~\cite{dagotto_94}. Even though experimental systems have a finite extent and open boundary conditions (OBCs), when disorder and imperfections can be neglected, theoretical studies typically employ simplified translationally invariant models with periodic boundary conditions (PBCs). In those models, quasimomentum is conserved so that, e.g., for noninteracting electrons one has a well-defined band structure that can be used to understand conventional metals and insulators. To take advantage of translational invariance, instead of studying local observables one generally studies their translationally invariant sums. This simplifies analytical derivations and enables numerical calculations in larger system sizes. Differences in boundary conditions, on the other hand, are important when one is interested in edge or surface effects, for example, in topological insulators~\cite{hasan_kane_10}, can result in subtleties when studying transport properties, e.g., the Drude weight~\cite{rigol_shastry_08,bellomia_resta_20}, and can change the scaling of finite-size effects in numerical calculations~\cite{iyer_15}. In this Letter we study the differences between local observables and their translationally invariant counterparts in the context of the eigenstate thermalization hypothesis (ETH) for clean systems with OBCs as well as with PBCs.

In generic isolated quantum systems, the equilibration of observables to their thermal equilibrium values is understood to be a consequence of eigenstate thermalization. The corresponding mathematical ansatz for the matrix elements $O^{}_{mn}\equiv\langle \psi^{}_m|\hat O| \psi^{}_n \rangle$ of an observable $\hat O$ in the eigenbasis $\{\ket{\psi^{}_m}\}$ of a quantum-chaotic Hamiltonian $\hat H$ ($\hat H\ket{\psi^{}_m}=E^{}_m\ket{\psi^{}_m}$) is known as the ETH~\cite{deutsch_1991, srednicki_1994, *srednicki_1999, rigol_2008, dalessio_quantum_2016},
\begin{equation}\label{eq:ETH}
 O^{}_{mn}=O(E^{}_m)\delta^{}_{mn}\,+\,\Omega^{-\frac{1}{2}}(\bar E^{}_{mn})\,\mathrm{f}^{}_O(\bar E^{}_{mn},\omega^{}_{mn})R^O_{mn}\,,
 \end{equation}
where $\bar E^{}_{mn}=(E^{}_m+E^{}_n)/2$ is the average energy, $\omega^{}_{mn}=E^{}_m-E^{}_n$ is the difference, $O(E^{}_m)$ and $\mathrm{f}^{}_O(\bar E^{}_{mn},\omega^{}_{mn})$ are smooth functions, $\Omega(\bar E^{}_{mn})$ is the density of states at the energy $\bar E^{}_{mn}$, and $R^O_{mn}$ are approximately normally distributed random numbers with zero mean and unit variance (variance 2) for $m\neq n$ ($m=n$) in systems with time-reversal symmetry. The ETH~\eqref{eq:ETH}, which characterizes the first two moments of the distribution of $O^{}_{mn}$, has been extended to include correlations needed to describe quantities involving higher moments~\cite{foini_kurchan_19, pappalardi_foini_kurchan_22}, such as the out-of-time-order correlators~\cite{foini_kurchan_19, chan_deluca_19, murthy_srednicki_19, brenes_pappalardi_21, wang_lamann_22, pappalardi_25}. It has also been proposed that correlations exist between matrix elements of different observables~\cite{dalessio_quantum_2016, foini_kurchan_19, noh_20, pappalardi_foini_kurchan_22}. Those correlations are central to our findings in this study.  

In the grand-canonical ensemble, the expectation value of an observable $\hat O$ at inverse temperature $\beta$ can be written as $O(\beta)=\tfrac1Z\text{Tr}(\hat O e^{-\beta\hat H})$, where $Z=\text{Tr}(e^{-\beta\hat H})$, and $\hat H$ is the Hamiltonian. The corresponding average energy is $E(\beta)=\tfrac1Z\text{Tr}(\hat H e^{-\beta\hat H})$. Using the ETH ansatz~\eqref{eq:ETH}, one finds that $O(\beta)\simeq O(E_m=E(\beta))$~\cite{dalessio_quantum_2016}. We are interested in eigenstate thermalization for local observables $\hat O^j$, at site $j$ of chains with $L$ sites, and their translationally invariant (intensive) counterparts 
\begin{equation}\label{eq:trans-invariant-observable}
    \hat \tio\equiv\frac1L\sum_{j=1}^L\hat O^j\,.
\end{equation}
In the presence of translational symmetry, $O^j_{mm}=O^l_{mm}\, \forall\, j,\, l$ so $\braket{\hat \tio^{}}^{}_\text{PBC}=\braket{\hat O^j}^{}_\text{PBC}$, where $\braket{\cdot}^{}_\text{PBC}$ stands for thermal averages in the presence of PBCs. On the other hand, for clean systems with OBCs, $\braket{\hat \tio^{}}^{}_\text{OBC}\simeq \braket{\hat O^j}^{}_\text{OBC}$ for $j$ in the bulk, and $\braket{\cdot}^{}_\text{OBC}$ are the corresponding averages in the presence of OBCs. Finally, $\braket{\hat \tio}^{}_\text{OBC}\simeq \braket{\hat \tio}^{}_\text{PBC}$, and their difference in general vanishes polynomially with $L$~\cite{iyer_15}. These facts motivate the interchangeable use of local observables and their translationally invariant counterparts, as well as of PBCs and OBCs.

\emph{Hamiltonians and diagonal ETH.} We study the \mbox{spin-1} tilted-field Ising model in chains with PBCs and OBCs, 
\begin{align}
    \hat H^{}_\text{PBC}&=J\sum_{j=1}^L \hat S_z^j\hat S_z^{j+1} +h^{}_x\sum_{j=1}^L\hat S_x^j +h^{}_z\sum_{j=1}^L\hat S_z^j\,, \label{eq:H_PBC}\\
    \hat H^{}_\text{OBC}&=\hat H^{}_\text{PBC} - J \hat S_z^L\hat S_z^{1} + \varepsilon \hat S_z^1\,, \label{eq:H_OBC}
\end{align}
where $\hat{S}_\gamma^j$ is the spin-1 operator along the $\gamma\in \{x,y,z\}$ direction at site $j$. We take $J=0.707$, $h^{}_x=1.1$, and $h^{}_z=0.9$ to study the nonintegrable regime~\cite{capizzi_poletti_24}. In chains with OBCs~\eqref{eq:H_OBC}, a weak field (of strength $\varepsilon=0.1$~\cite{Supp_Mat}) is added in the first site along $z$ to break the lattice reflection symmetry. We focus on the local observable $\hat O^j= \hat S_x^j$ at site $j=\lceil L/2\rceil$ and its translationally invariant counterpart $\hat \tio=\tfrac1L\sum_{j=1}^L \hat S_x^j$~\cite{Supp_Mat}.

\begin{figure}[!t]
    \includegraphics[width=\columnwidth]{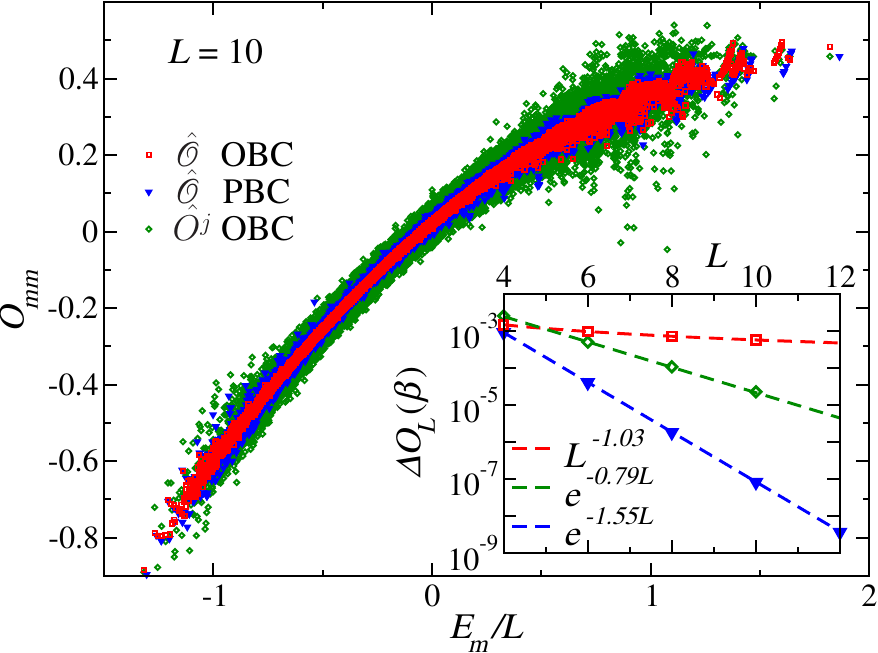}
    \vspace{-0.55cm}
    \caption{Diagonal matrix elements $O^{}_{mm}$ vs the energy density $E^{}_m/L$ for the local observable $\hat O^j=\hat S_x^j$ at site $j=\tfrac L2$ for OBCs, and its translationally invariant counterpart $\hat \tio=\tfrac1L\sum_j\hat S_x^j$ for OBCs and PBCs, in chains with $L=10$. Inset: Relative difference $\Delta O^{}_{L}(\beta)$ between $O^{}_{L}(\beta)$ in a chain with $L$ sites and its thermodynamic limit value $O^{}_\text{tl}(\beta)$ for $\beta=0.5$ (see text). $O^{}_\text{tl}(\beta)$ is obtained by fitting the PBC results for the largest four chains to $O^{}_{L}(\beta)=O^{}_\text{tl}(\beta)+ae^{-bL}$ with fitting parameters $O^{}_\text{tl}(\beta),\ a$, and $b$. This value is then used for OBCs to fit $\Delta O^{}_{L}(\beta)$ for $\hat O^j$ to $ae^{-bL}$, and for $\hat \tio$ to $aL^{-b}$, with fitting parameters $a$ and $b$. The fits are shown as lines.}
    \label{fig:diagonal-elements}
\end{figure}

In Fig.~\ref{fig:diagonal-elements} we plot the diagonal matrix elements $O^{}_{mm}$ versus $E^{}_m/L$ for $\hat O^j=\hat S_x^j$ and $\hat\tio=\tfrac1L\sum_j\hat S_x^j$ in chains with OBCs, and for $\hat \tio$ in chains with PBCs. The results are consistent with the ETH in all three cases, but the magnitudes of the eigenstate-to-eigenstate fluctuations depend on the observable and the boundary conditions. Those fluctuations are similar for $\hat\tio$ independently of the boundary conditions, are largest for $\hat O^j$ in chains with OBCs, and we verified that they decrease exponentially with system size in all cases (not shown). To quantify how the grand-canonical expectation values in finite chains $O^{}_{L}(\beta)$ converge to the thermodynamic limit result $O^{}_\text{tl}(\beta)$, we compute the relative difference $\Delta O^{}_{L}(\beta)=|O^{}_{L}(\beta)-O^{}_\text{tl}(\beta)|/|O^{}_\text{tl}(\beta)|$. As shown in the inset in Fig.~\ref{fig:diagonal-elements}, the relative difference for $\hat\tio$ decreases exponentially fast for PBCs and polynomially fast for OBCs~\cite{iyer_15}. Remarkably, for $\hat O^j$ with $j$ at the center of the chain, the decrease is exponentially fast for OBCs. This indicates that, so long as one is interested in the bulk expectation values of local observables, it does not matter (up to finite-size effects) whether one computes the local observables or their translationally invariant counterparts in chains with OBCs or PBCs.

\emph{Spectral functions.}
We study the off-diagonal matrix elements through the spectral functions at infinite temperature, i.e., for $E^{}_\infty\equiv\tfrac1D\text{Tr}(\hat H)$, where $D=3^L$ is the Hilbert space dimension ($E^{}_\infty=0$ for our Hamiltonians). For $\hat O=\hat O^j$ and $\hat\tio$~\eqref{eq:trans-invariant-observable}, the spectral functions read
\begin{eqnarray}\label{eq:spectral-function_local}
    |f^{}_{O^j}(E^{}_\infty,\omega)|^2=\frac{1}{D}\sum_{m,n\neq m}|O^j_{mn}|^2\delta(\omega-\omega^{}_{mn})\,,\\
    |f^{}_{\tio}(E^{}_\infty,\omega)|^2=\frac{L}{D}\sum_{m,n\neq m}|\tio_{mn}|^2\delta(\omega-\omega^{}_{mn})\,, \label{eq:spectral-function_transl}
\end{eqnarray}
see, e.g., Refs.~\cite{pandey_claeys_20, leblond_sels_21, kim_polkovnikov_24, capizzi_poletti_24, abdelshaf_mondaini_25}, where the factor of $L$ in Eq.~\eqref{eq:spectral-function_transl} is a consequence of the Hilbert-Schmidt norm of $\hat \tio$~\cite{mierzejewski_vidmar_20, patrycja_rafal_24}. We use the regularization $\delta(x)\rightarrow \frac{1}{\sqrt{2\pi \sigma^2}}\exp(-\frac{x^2}{2\sigma^2})$~\cite{abdelshaf_mondaini_25}, with $\sigma=0.1\omega^{}_H$, where $\omega^{}_H$ is the average level spacing in the central $50\%$ of the spectrum.

\begin{figure}[!t]
    \includegraphics[width=\columnwidth]{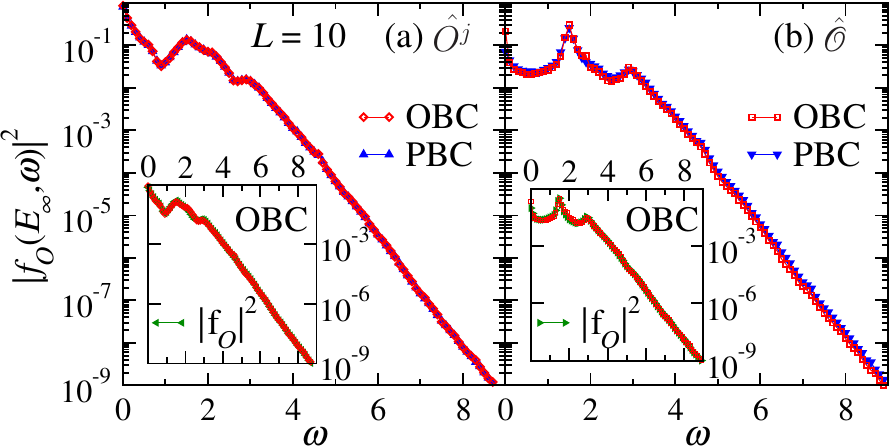}
    \vspace{-0.5cm}
    \caption{Spectral functions of (a) $\hat O^j=\hat S_x^j$ at site $j=L/2$, and (b) its translationally invariant counterpart $\hat \tio=\tfrac1L\sum_j \hat S_x^j$ in chains with PBCs and OBCs, with $L=10$. Insets: Spectral function $|f^{}_O(E^{}_\infty,\omega)|^2$ (same as in the main panel) and the rescaled ETH function $|\mathrm{f}^{}_O(\bar E^{}_{mn}\approx E^{}_\infty,\omega)|^2 \exp\left(-\frac{\omega^2}{4\sigma^2_E}\right)$ in chains with OBCs. $|\mathrm{f}^{}_O(\bar E^{}_{mn}\approx E^{}_\infty,\omega)|^2$ is obtained by calculating the variance of the off-diagonal matrix elements $\overline{|O^{}_{mn}|^2}$ [see Eq.~\eqref{eq:ETH}] using a running average with a window of size $\Delta\omega=0.15$, evaluated every $\delta\omega=0.1$, with $\bar E^{}_{mn}$ restricted to the central $10\%$ of the energy spectrum.}
    \label{fig:both-boundary-conditions}
\end{figure}

The spectral functions $|f^{}_{O}(E^{}_\infty,\omega)|^2$, which are well defined even if the ETH does not apply, e.g., in integrable systems~\cite{leblond_2019, leblond_sels_21}, converge to $|\mathrm{f}^{}_{O}(E^{}_\infty,\omega)|^2$ in the ETH ansatz~\eqref{eq:ETH} in the thermodynamic limit. In finite systems, however, the two are related by $|f^{}_O(E^{}_\infty,\omega)|^2\simeq\exp\left(-\frac{\omega^2}{4\sigma_E^2}\right) |\mathrm{f}^{}_O(E^{}_\infty,\omega)|^2$, where $\sigma^2_E$ is the energy variance~\cite{Supp_Mat, patil_rigol_ETH_review_26}. This is shown by the data collapse in the insets in Fig.~\ref{fig:both-boundary-conditions}, in which we plot numerical results obtained independently for $|f^{}_O(E^{}_\infty,\omega)|^2$ and $\exp\left(-\frac{\omega^2}{4\sigma_E^2}\right) |\mathrm{f}^{}_O(E^{}_\infty,\omega)|^2$.

\begin{figure}[!t]
    \includegraphics[width=\columnwidth]{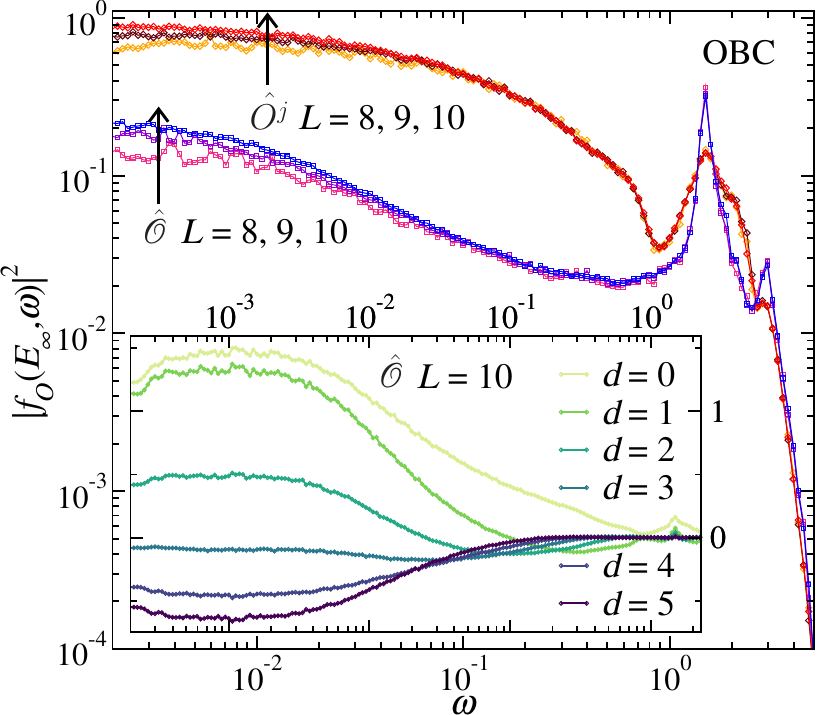}
    \vspace{-0.5cm}
    \caption{Spectral functions of the local observable $\hat O^j=\hat S_x^j$ at site $j=\lceil L/2\rceil$ and its translationally invariant counterpart $\hat \tio=\tfrac1L\sum_j \hat S_x^j$ in chains with OBCs~\eqref{eq:H_OBC}. Inset: Contributions to $|f^{}_{\tio}(E^{}_\infty,\omega)|^2$ from the terms in Eq.~\eqref{eq:termsinTI} with \mbox{$d\equiv |j-l|=0$\,--\,5} for $L=10$.}
    \label{fig:open-chain}
\end{figure}

{\bf OBCs}. We first study chains with OBCs, for which the Hamiltonian has no trivial lattice symmetries. In Fig.~\ref{fig:open-chain}, we show $|f^{}_{O^j}(E^{}_\infty,\omega)|^2$ and $|f^{}_{\tio}(E^{}_\infty,\omega)|^2$. The two are quantitatively different but exhibit qualitatively similar behaviors: (i) at low frequencies they both develop the plateaus expected in nonintegrable systems below the Thouless energy~\cite{dalessio_quantum_2016}, and (ii) at high frequencies they exhibit a fast decay. To understand the origin of their differences, we rewrite $|\tio_{mn}|^2$ as
\begin{equation}
  |\tio_{mn}|^2=\frac{1}{L^2}\bigg[\sum_{j=1}^L |O^j_{mn}|^2+ {\sum_{j,\,l=1\, (j\neq l)}^L} O^j_{mn} (O^l_{mn})^*\bigg]. \label{eq:termsinTI} 
\end{equation}
Hence, if $R^{O^j}_{mn}$ and $R^{O^l}_{mn}$ in Eq.~\eqref{eq:ETH} for $j\neq l$ are taken to be uncorrelated (such as in random matrix theory~\cite{dalessio_quantum_2016}), then for $m\neq n$ the average $\overline{O^j_{mn} (O^l_{mn})^*}=0$. Also, if $\overline{|O^j_{mn}|^2}$ varies little throughout the bulk of the system, then $\overline{|\tio_{mn}|^2}\approx \tfrac{1}{L}\overline{|O^j_{mn}|^2}$. In this case the two spectral functions are approximately the same. The differences between $|f^{}_{O^j}(E^{}_\infty,\omega)|^2$ and $|f^{}_{\tio}(E^{}_\infty,\omega)|^2$ in clean systems with OBCs therefore signal the existence of correlations between the matrix elements of $\hat O^j$ at different sites that are beyond the traditional random matrix theory formulation of the ETH. This is explicitly shown in the inset in Fig.~\ref{fig:open-chain}, where we plot the contributions to $|f^{}_{\tio}(E^{}_\infty,\omega)|^2$ from terms with different values of $d\equiv|j-l|$ in Eq.~\eqref{eq:termsinTI}, where $d=0, 1,\ldots,L-1$. Those contributions depend on $d$ and can be negative for $d>0$. The relative ordering between $|f^{}_{\tio}(E^{}_\infty,\omega)|^2$ and $|f^{}_{O^j}(E^{}_\infty,\omega)|^2$ is observable-dependent. For some observables, such as $\hat S_x$ considered here, those correlations result in $|f^{}_{\tio}(E^{}_\infty,\omega)|^2<|f^{}_{O^j}(E^{}_\infty,\omega)|^2$ at low frequencies, while for others $|f^{}_{\tio}(E^{}_\infty,\omega)|^2>|f^{}_{O^j}(E^{}_\infty,\omega)|^2$~\cite{Supp_Mat}. A different perspective on these findings, supporting the expectation that the spectral functions $|f^{}_{\tio}(E^{}_\infty,\omega)|^2$ and $|f^{}_{O^j}(E^{}_\infty,\omega)|^2$ are in general different in local models, is provided by systems with PBCs, as we explain next.

\begin{figure}[!t]
    \includegraphics[width=\columnwidth]{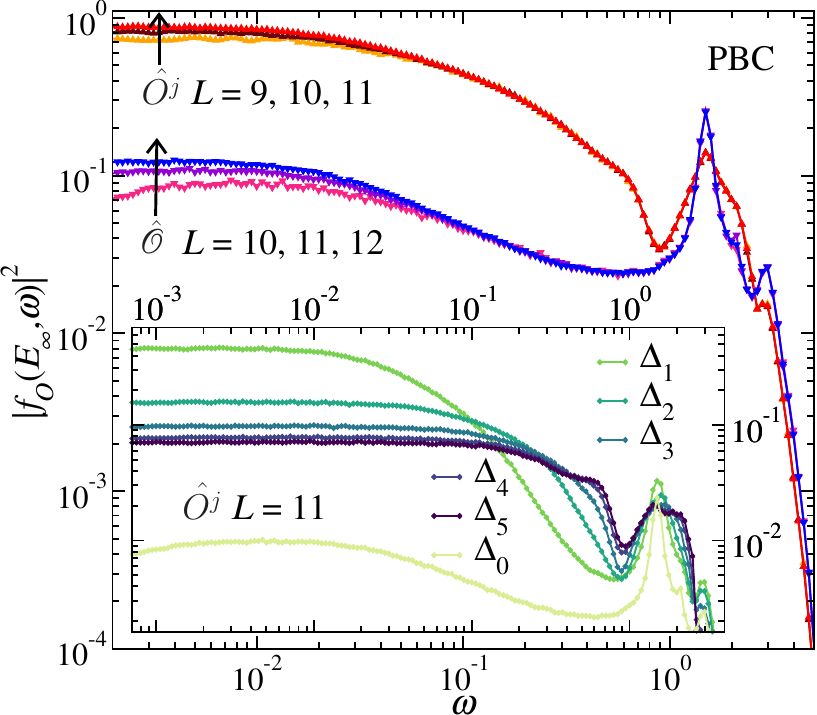}
    \vspace{-0.5cm}
    \caption{Spectral functions of the local $\hat O^j=\hat S_x^j$ and its translationally invariant counterpart $\hat \tio=\tfrac1L\sum_j \hat S_x^j$ in chains with PBCs~\eqref{eq:H_PBC}. Inset: Contributions to $|f^{}_{O^j}(E^{}_\infty,\omega)|^2$ from the six terms in Eq.~\eqref{eq:PBC-local-spectral-function} with different values of $\Delta^{}_\ell$ for $L=11$.}
    \label{fig:periodic-chain}
\end{figure}

{\bf PBCs.} We begin our discussion of systems with PBCs by stressing that, being well-defined physical quantities, the spectral functions of local observables and their translationally invariant counterparts are expected to be (up to finite-size effects) independent of the boundary conditions~\cite{rigol_shastry_08}. In Fig.~\ref{fig:both-boundary-conditions}, we show results for $|f^{}_{O^j}(E^{}_\infty,\omega)|^2$ [Fig.~\ref{fig:both-boundary-conditions}(a)] and $|f^{}_{\tio}(E^{}_\infty,\omega)|^2$ [Fig.~\ref{fig:both-boundary-conditions}(b)] in chains with PBCs and OBCs for the same number of lattice sites ($L=10$). The results for $|f^{}_{O^j}(E^{}_\infty,\omega)|^2$ are indistinguishable between OBCs and PBCs [Fig.~\ref{fig:both-boundary-conditions}(a)], while small differences are visible for $|f^{}_{\tio}(E^{}_\infty,\omega)|^2$ [Fig.~\ref{fig:both-boundary-conditions}(b)] because of the finite-size boundary effects. The log-linear scale used makes apparent that the spectral functions decay exponentially fast with $\omega$~\cite{leblond_2019}. In Fig.~\ref{fig:periodic-chain}, we plot the spectral functions of $\hat O^j$ and $\hat \tio$ for the three largest chains with PBCs~\eqref{eq:H_PBC} that we diagonalized on a log-log scale to emphasize the low-frequency behavior. As expected, they closely resemble the OBC results shown in Fig.~\ref{fig:open-chain}. 

In translationally invariant chains~\eqref{eq:H_PBC}, the differences between $|f^{}_{\tio}(E^{}_\infty,\omega)|^2$ and $|f^{}_{O^j}(E^{}_\infty,\omega)|^2$ can be understood as follows. The eigenstates of $\hat H^{}_\text{PBC}$~\eqref{eq:H_PBC} are simultaneous eigenstates of the single-site lattice translation operator $\hat T$, i.e., $\hat H^{}_\text{PBC}\ket{E^{}_m,k^{}_m}=E^{}_m\ket{E^{}_m,k^{}_m}$ and $\hat T\ket{E^{}_m,k^{}_m}=e^{-ik^{}_m}\ket{E^{}_m,k^{}_m}$, with energy $E^{}_m$ and quasimomentum $k^{}_m=\tfrac{2\pi}{L}\eta^{}_m$ (we set $\hbar=1$ and the lattice spacing $a=1$), where $\eta^{}_m=\eta^{}_\text{min},\eta^{}_\text{min}+1,\ldots,\eta^{}_\text{max}$ with $\eta^{}_\text{max}=-\eta^{}_\text{min}+1=\tfrac{L}{2}$ ($\eta^{}_\text{max}=-\eta^{}_\text{min}=\lfloor \tfrac L2\rfloor$) for $L$ even (odd). This means that, for any observable $\hat O$, the matrix $O^{}_{mn}\equiv\bra{E^{}_m,k^{}_m}\hat O\ket{E^{}_n,k^{}_n}$ decomposes into $L^2$ {\it blocks} of about the same size labeled by the values of $k^{}_m$ and $k^{}_n$. Since local operators at different sites in a periodic chain are related by a translation, $\hat O^l=(\hat T^\dagger)^{(j-l)} \hat O^j(\hat T)^{(j-l)}$, we find that
\begin{align}\label{eq:different-sites-phase-relation}
    O^l_{mn}=O^j_{mn}e^{i(j-l)(k^{}_m-k^{}_n)}\,.
\end{align}
This implies that (i) for $k^{}_m=k^{}_n$, $O^j_{mn}=O^l_{mn}\ \forall\, j,\, l$ so that $\tio_{mn}=O^j_{mn}\ \forall\, j$, and (ii) $\tio_{mn}=0$ for $k^{}_m \neq k^{}_n$.

Consequently, for the spectral function of translationally invariant operators $\hat \tio$ in Eq.~\eqref{eq:spectral-function_transl}, the translational symmetry restricts the sums over $n$ and $m$ to pairs of energy eigenstates with the same quasimomentum:
\begin{equation}\label{eq:PBC-trans-spectral-function}
    |f^{}_{\tio}(E^{}_\infty,\omega)|^2=\frac{L}{D}\sum_{m,\,n\neq m\, (k^{}_m=k^{}_n)}\! |\tio_{mn}|^2\delta(\omega-\omega^{}_{mn})\,.
\end{equation}
The spectral function for local operators, on the other hand, also has contributions from pairs of energy eigenstates with different quasimomenta. Motivated by Eq.~\eqref{eq:PBC-trans-spectral-function}, we rewrite Eq.~\eqref{eq:spectral-function_local} by grouping all contributions from energy eigenstates with the same value of $|k^{}_m-k^{}_n|=\Delta^{}_\ell$, with $\Delta^{}_\ell=\tfrac{2\pi}{L}\ell$ and $\ell =0,\ 1,\ \ldots,\ \lfloor\frac{L}{2}\rfloor$:
\begin{equation}\label{eq:PBC-local-spectral-function}
    |f^{}_{O^j}(E^{}_\infty,\omega)|^2=\frac{1}{D}\sum_{\ell=0}^{\lfloor\frac{L}{2}\rfloor} \sum_{\substack{m,\, n\neq m\\ (k_n=k_m\pm\Delta_\ell)}} |O^j_{mn}|^2\delta(\omega-\omega^{}_{mn}).
\end{equation}
The terms in Eq.~\eqref{eq:PBC-local-spectral-function} with $\Delta^{}_\ell=0$ add to $\tfrac{1}{L}|f^{}_{\tio}(E^{}_\infty,\omega)|^2$, because of the prefactor $L$ in Eq.~\eqref{eq:PBC-trans-spectral-function}. This shows that their contribution to $|f^{}_{O^j}(E^{}_\infty,\omega)|^2$ vanishes as $\tfrac{1}{L}$. Consequently, in the thermodynamic limit, $|f^{}_{O^j}(E^{}_\infty,\omega)|^2$ is determined by the terms with $\Delta^{}_\ell\neq 0$. Since there is no symmetry requirement for those terms to add to $L-1$ times the $\Delta^{}_\ell=0$ contribution, we conclude that, in general, $|f^{}_{\tio}(E^{}_\infty,\omega)|^2$ and $|f^{}_{O^j}(E^{}_\infty,\omega)|^2$ are different in translationally invariant systems. This, together with the equivalence of the spectral functions in systems with PBCs and OBCs, implies that the correlations observed between the matrix elements of local operators at different sites in systems with OBCs are generic.

In the inset of Fig.~\ref{fig:periodic-chain}, we plot the contributions from the six values of $\Delta^{}_\ell$ in Eq.~\eqref{eq:PBC-local-spectral-function} to the spectral function of $\hat O^j$ for $L=11$. As anticipated, they depend on the value of $\Delta^{}_\ell$. For $\Delta^{}_\ell\neq 0$, they exhibit a monotonic behavior with increasing $\Delta^{}_\ell$. The results for $\Delta^{}_1$--$\Delta^{}_5$, from top to bottom, show that the onset of the low-frequency plateau moves to higher frequencies as $\Delta^{}_\ell$ increases, which reflects that the longest time scales in the dynamics of $\hat O^j$ associated with a given $\Delta^{}_\ell$ decrease with increasing $\Delta^{}_\ell$, as expected from a decreasing length scale $\propto\tfrac{1}{\Delta^{}_\ell}$. The contribution with $\Delta^{}_\ell=0$ is special because it is not associated with a finite length scale. At low frequencies, this results in differences between $|f^{}_{\tio}(E^{}_\infty,\omega)|^2$ in systems with PBCs and OBCs. For OBCs, the system size $L$ provides the longest length scale and $|f^{}_{\tio}(E^{}_\infty,\omega)|^2$ is similar at low frequency to the curve for $\Delta^{}_{\ell=1}=\tfrac{2\pi}{L}$ in the inset in Fig.~\ref{fig:periodic-chain}, which corresponds to the contribution from the longest finite length scale for PBCs.

Our results in the inset in Fig.~\ref{fig:periodic-chain}, together with the fact that local operators have approximately normally distributed matrix elements for $k^{}_m \neq k^{}_n$~\cite{leblond_2020}, raise the question of whether there exists a formulation of the ETH in translationally invariant systems beyond Eq.~\eqref{eq:ETH} for translationally invariant operators within each quasimomentum sector~\cite{rigol_09a,*rigol_09b}. Generalizations of the ETH have been recently explored in the context of continuous non-Abelian symmetries~\cite{murthy_nonabelian_2023, noh_2023, lasek_24, patil_rigol_25}. In contrast to continuous symmetries, lattice-translational symmetry is discrete so it does not have an associated extensive generator.

Remarkably, despite the special nature of lattice translational symmetry, we find that there is a formulation of the ETH that holds in periodic chains. It reads
\begin{align}\label{eq:ETH_TI}
\bra{E^{}_m,k^{}_m}\hat O&\ket{E^{}_n,k^{}_n} = O(E^{}_m)\delta^{}_{mn}\\& + \Omega^{-\frac{1}{2}}(\bar E^{}_{mn})\,\mathrm{f}^{}_O(\bar E^{}_{mn},\omega^{}_{mn},\kappa^{}_{mn})R^O_{mn}\,,\nonumber
 \end{align}
where, in addition to the quantities entering Eq.~\eqref{eq:ETH}, it also depends on $\kappa^{}_{mn}=|k^{}_m-k^{}_n|$. Notable in Eq.~\eqref{eq:ETH_TI} is the lack of dependence of $O(E^{}_m)$ on $k^{}_m$~\cite{rigol_09a,*rigol_09b} and the lack of dependence of the density of states on the quasimomenta involved (because all quasimomentum blocks asymptotically have the same density of states). Furthermore, $\mathrm{f}^{}_O$ is independent of the average quasimomentum $(k^{}_m+k^{}_n)/2$, and the quasimomenta only enter $\mathrm{f}^{}_O$ through the quasimomentum difference $\kappa^{}_{mn}=|k^{}_m-k^{}_n|$.

\begin{figure}[!t]
    \includegraphics[width=\columnwidth]{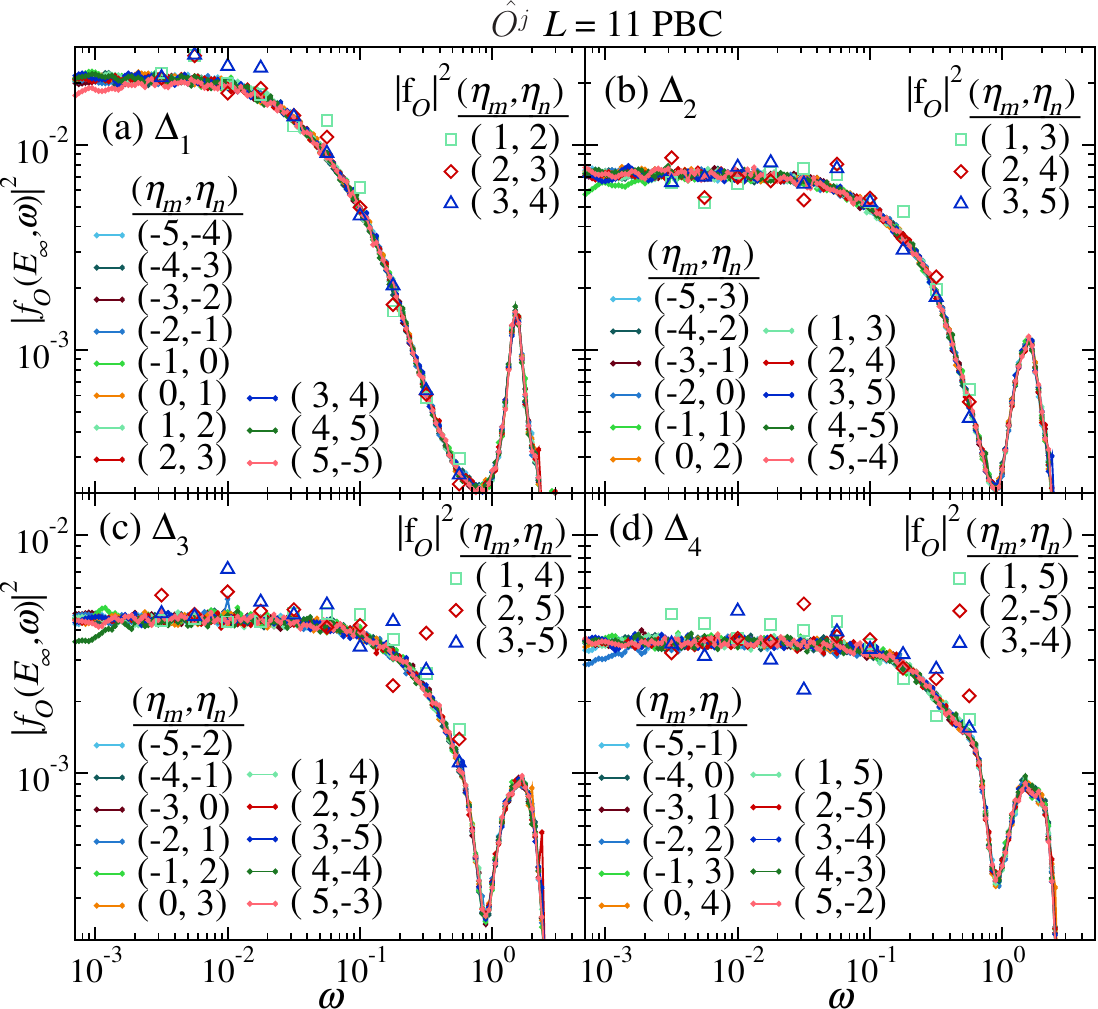}
    \vspace{-0.5cm}
    \caption{Contributions from quasimomentum blocks (solid lines) labeled by $(k^{}_m,k^{}_n)=\tfrac{2\pi}{L}(\eta^{}_m,\eta^{}_n)$ to the results shown in the inset in Fig.~\ref{fig:periodic-chain} for different values of $\Delta^{}_\ell$ with $\ell=1,\,2,\,3,\,4\,$ in (a)--(d), respectively, for $L=11$. Equation~\eqref{eq:PBC-local-spectral-function} shows how those terms enter in $|f^{}_{O^j}(E^{}_\infty,\omega)|^2$. Open symbols show the rescaled ETH functions $\tfrac{1}{L^2}|\mathrm{f}^{}_{O^j}|^2$ in blocks labeled by $(k^{}_m,k^{}_n)=\tfrac{2\pi}{L}(\eta^{}_m,\eta^{}_n)$ with $\eta^{}_m=1,\,2,\,3$ and $k^{}_n=k^{}_m+\Delta^{}_\ell$, calculated using a coarse-grained (frequency-resolved) variance of the off-diagonal elements $\overline{|O^j_{mn}|^2}$ over 100 pairs of states with $|\bar E^{}_{mn}- E^{}_\infty|\leqslant0.2$ and $|\log_{10}\omega^{}_{mn}-\log_{10}\omega|\leqslant0.01$ for values of $\omega$ separated by $\delta(\log_{10}\omega)=0.25$.}
    \label{fig:PBC-quasimomentum-resolved}
\end{figure} 

We test Eq.~\eqref{eq:ETH_TI} by taking a closer look at the $\Delta^{}_\ell$ curves in the inset in Fig.~\ref{fig:periodic-chain} and decomposing them into contributions from the individual quasimomentum blocks labeled by $(k^{}_m,k^{}_n)=(k^{}_m,k^{}_m+\Delta^{}_\ell)$, as shown in Fig.~\ref{fig:PBC-quasimomentum-resolved}. The collapse of curves for different blocks contributing to a given $\Delta^{}_\ell$ indicates the absence of a finer structure at the level of the specific $(k^{}_m,k^{}_n)$ blocks, and $\kappa^{}_{mn}=|k^{}_m-k^{}_n|=\Delta^{}_\ell$ emerges as the relevant quantity characterizing the terms in $|f^{}_{O^j}(E^{}_\infty,\omega)|^2$, consistent with the formulation of the ETH in Eq.~\eqref{eq:ETH_TI}.

To probe Eq.~\eqref{eq:ETH_TI} to an ever finer degree, we calculate the ETH function $|\mathrm{f}^{}_{O^j}(\bar E^{}_{mn},\omega^{}_{mn},\kappa^{}_{mn})|^2$ for $\kappa^{}_{mn}=\Delta^{}_\ell$ and $\bar E^{}_{mn}\approx E^{}_\infty$ as
\vspace{-0.2cm}
\begin{equation}
    |\mathrm{f}^{}_{O^j}(E^{}_\infty,\omega,\Delta^{}_\ell)|^2=\Omega(E_\infty)\,\overline{|O^j_{mn}|^2}\,,
\end{equation}
\vspace{-0.4cm}

\noindent where $\Omega(E_\infty)=D/\sqrt{2\pi\sigma_E^2}$ is the density of states at the center of the spectrum, with energy variance $\sigma^2_E$, and $\overline{|O^j_{mn}|^2}\equiv\overline{|O^j_{mn}|^2}(\omega)$ is the coarse-grained frequency-resolved variance of the off-diagonal matrix elements $O^j_{mn}$, calculated using 100 pairs of energy eigenstates in the block labeled by $(k^{}_m,k^{}_n)$ with $\kappa^{}_{mn}=|k^{}_m-k^{}_n|=\Delta^{}_\ell$, energy $|\bar E^{}_{mn}- E^{}_\infty|\leqslant0.2$ and frequency $|\log_{10}\omega^{}_{mn}-\log_{10}\omega|\leqslant0.01$ for various values of $\omega$. From the pairs of energy eigenstates satisfying the above constraints, we select the 100 pairs with the smallest $\max(|E^{}_m-E^{}_\infty|,|E^{}_n-E^{}_\infty|)$ to calculate $\overline{|O^j_{mn}|^2}$.

In Fig.~\ref{fig:PBC-quasimomentum-resolved}, we show the results for the ETH functions $\tfrac{1}{L^2}|\mathrm{f}^{}_{O^j}(E^{}_\infty,\omega,\Delta^{}_\ell)|^2$ (open symbols), rescaled by $\tfrac{1}{L^2}$ to compare them  with the contributions to the spectral function $|f^{}_{O^j}(E^{}_\infty,\omega)|^2$ in Eq.~\eqref{eq:PBC-local-spectral-function} from fixed $(k^{}_m,k^{}_n)$ blocks, which vanish as $\tfrac{1}{L^2}$. Remarkably, the rescaled ETH functions $\tfrac{1}{L^2}|\mathrm{f}^{}_{O^j}(E^{}_\infty,\omega,\Delta^{}_\ell)|^2$ (obtained using 100 pairs of energy eigenstates) agree with the corresponding contributions to $|f^{}_{O^j}(E^{}_\infty,\omega)|^2$ computed using all matrix elements within a quasimomentum block $(k^{}_m,k^{}_m+\Delta^{}_\ell)$.

\emph{Summary}. We explored the ETH for local and translationally invariant operators in clean chains with OBCs and PBCs. For nonintegrable chains with OBCs~\eqref{eq:H_OBC} in the absence of symmetries, where the ETH applies to the entire energy spectrum, the spectral functions for the local operator $\hat S_x^j$ at the center of the chains and its translationally invariant counterpart were found to exhibit differences, which we traced back to correlations between the off-diagonal matrix elements of the local operators $\hat S_x^j$ at different sites. For translationally invariant chains with PBCs~\eqref{eq:H_PBC}, in which the matrix elements of local operators at different sites are related by translation symmetry, we argued that the spectral functions of local and translationally invariant operators should in general be different. Our findings highlight that in local models the spectral functions of local and translationally invariant operators cannot, in general, be used interchangeably in clean systems, independently of the boundary conditions. 

By exploring the quasimomentum dependence of the matrix elements of $\hat S_x^j$, we unveiled an unexpected (due to the absence of an associated extensive generator) extension of the ETH for translationally invariant lattice systems~\eqref{eq:ETH_TI}. This extension needs to be explored further and opens the door to simplifying numerical calculations of spectral functions of local operators in translationally invariant systems by requiring calculations for only $\lfloor\frac{L}{2}\rfloor+1$ of the $L^2$ quasimomentum blocks.

\emph{Acknowledgments}.
This work was supported by the National Science Foundation (NSF) Grant No.~PHY-2309146. The computations were done in the ICDS Roar supercomputer at Penn State. 

\bibliography{references}

\clearpage

%%%%%%%%%%%%%%%%%%    Supplementary    %%%%%%%%%%%%%%%%%%%%%%
%\newpage
\phantom{a}
%\newpage
\setcounter{figure}{0}
\setcounter{equation}{0}

\renewcommand{\thetable}{S\arabic{table}}
\renewcommand{\thefigure}{S\arabic{figure}}
\renewcommand{\theequation}{S\arabic{equation}}
\renewcommand{\thesection}{S\arabic{section}}

\onecolumngrid
\begin{center}
{\large \bf Supplemental Material: Eigenstate~thermalization~for~local~versus~translationally~invariant~observables}\\

\vspace{0.2cm}

\begin{center}
Rohit Patil and Marcos Rigol\\
{\it Department of Physics, The Pennsylvania State University, University Park, Pennsylvania 16802, USA}
\end{center}

\end{center}

\vspace{0.7cm}
\twocolumngrid

\section{Spectral functions for nearest-neighbor correlator}\label{sec:S1}

Here, we provide additional evidence for the difference between the spectral functions of local observables and their translationally invariant counterparts. We consider as the local observable the nearest-neighbor correlator $\hat O^j=\hat S_x^j\hat S_x^{j+1}$ with $j=\lceil\tfrac{L}{2}\rceil$ and its translationally invariant counterpart $\hat\tio=\tfrac1N\sum_{j=1}^{N} \hat S_x^j\hat S_x^{j+1}$, where $N=L$ ($N=L-1$) in chains with PBCs (OBCs).

\begin{figure}[!t]
    \includegraphics[width=\columnwidth]{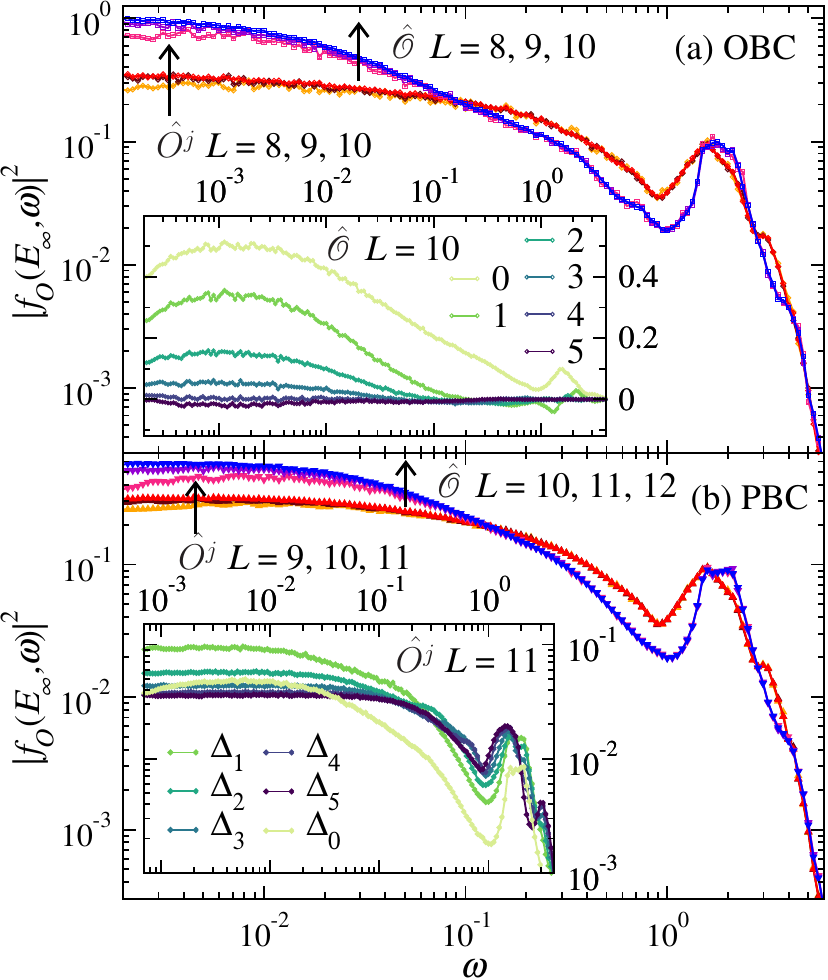}
    \vspace{-0.4cm}
    \caption{Spectral functions for the local observable $\hat O^j=\hat S_x^j\hat S_x^{j+1}$ with $j=\lceil\tfrac{L}{2}\rceil$ and its translationally invariant counterpart $\hat \tio=\tfrac1N\sum_{j=1}^N \hat S_x^j\hat S_x^{j+1}$ in chains with: (a) OBCs [see Eq.~\eqref{eq:H_OBC}] for which $N=L-1$, and (b) PBCs [see Eq.~\eqref{eq:H_PBC}] for which $N=L$. Inset in (a): Contributions to $|f^{}_{\tio}(E^{}_\infty,\omega)|^2$, see Eq.~\eqref{eq:termsinTI} where one needs to change $L\rightarrow L-1$, from terms with $d\equiv |j-l|=0$ through 5 in chains with $L=10$. Inset in (b): Contributions to $|f^{}_{O^j}(E^{}_\infty,\omega)|^2$ from different values of $\Delta^{}_\ell$ in Eq.~\eqref{eq:PBC-local-spectral-function} for $L=11$.}
    \label{fig:Sup-SxSx}
\end{figure}

In Fig.~\ref{fig:Sup-SxSx}, we plot the spectral functions $|f^{}_{\tio}(E^{}_\infty,\omega)|^2$ and $|f^{}_{O^j}(E^{}_\infty,\omega)|^2$ for the nearest-neighbor correlator in chains with OBCs [Fig.~\ref{fig:Sup-SxSx}(a)] and PBCs [Fig.~\ref{fig:Sup-SxSx}(b)], respectively. As for the $x$-magnetization in the main text, the spectral functions $|f^{}_{\tio}(E^{}_\infty,\omega)|^2$ and $|f^{}_{O^j}(E^{}_\infty,\omega)|^2$ are different for the nearest-neighbor correlator, independently of the boundary conditions. In contrast to the relative ordering $|f^{}_{\tio}(E^{}_\infty,\omega)|^2<|f^{}_{O^j}(E^{}_\infty,\omega)|^2$ observed at low frequencies for the $x$-magnetization in the main text, we find that $|f^{}_{\tio}(E^{}_\infty,\omega)|^2>|f^{}_{O^j}(E^{}_\infty,\omega)|^2$ for the nearest-neighbor correlator. These results support the statement made in the main text that the relative ordering between $|f^{}_{\tio}(E^{}_\infty,\omega)|^2$ and $|f^{}_{O^j}(E^{}_\infty,\omega)|^2$ in general depends on the observable.

In chains with OBCs, the differences between $|f^{}_{\tio}(E^{}_\infty,\omega)|^2$ and $|f^{}_{O^j}(E^{}_\infty,\omega)|^2$ can be attributed to the existence of correlations between $\hat O^j$ at different sites. These correlations can be seen in the inset in Fig.~\ref{fig:Sup-SxSx}(a), where we show contributions of different terms in Eq.~\eqref{eq:termsinTI} for $\hat \tio$ as a function of distance $d\equiv |j-l|$ between the sites $j,l$ involved. The nonvanishing results for the terms with $j\neq l$ ($d>0$) show the existence of correlations between the matrix elements of $\hat O^j$ at different sites.

In chains with PBCs, on the other hand, the differences between $|f^{}_{\tio}(E^{}_\infty,\omega)|^2$ and $|f^{}_{O^j}(E^{}_\infty,\omega)|^2$ can be understood by considering the different terms with $|k^{}_m-k^{}_n|=\Delta^{}_\ell$ in Eq.~\eqref{eq:PBC-local-spectral-function} for the local observable $\hat O^j$, which breaks the translational invariance of the model and connects eigenstates with different quasimomenta $k^{}_m,\,k^{}_n$. In the inset in Fig.~\ref{fig:Sup-SxSx}(b), we show the contributions for $\Delta^{}_0$ through $\Delta^{}_5$ for the local nearest-neighbor correlator $\hat O^j$ considered here. Similar to the results in the main text, the onset of low-frequency plateau moves towards higher frequencies with increasing $\Delta^{}_{\ell\neq0}$.  

\section{Effect of boundary term in chains with OBC}

\begin{figure}[!t]
    \includegraphics[width=\columnwidth]{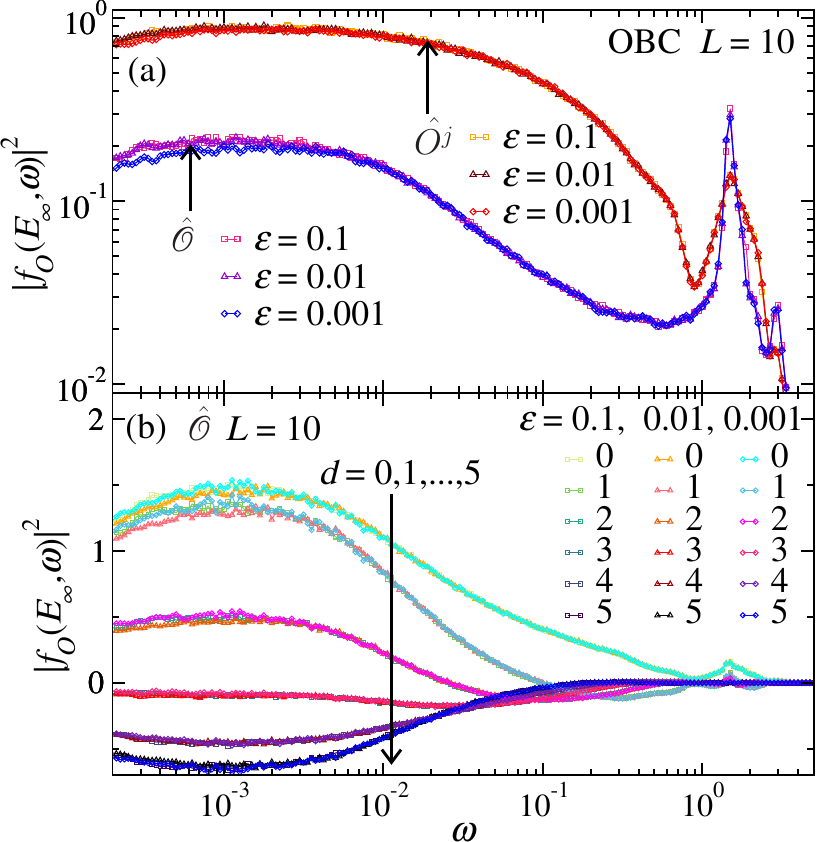}
    \vspace{-0.4cm}
    \caption{Comparison between OBC results for various values of the boundary field strength, $\varepsilon=0.1,0.01$ and $0.001$. (a) Spectral functions for the local observable $\hat O^j=\hat S_x^j$ at site $j=\lceil L/2\rceil$ and its translationally invariant counterpart $\hat \tio=\tfrac1L\sum_j \hat S_x^j$ in chains with OBCs~\eqref{eq:H_OBC}. (b) Contributions to $|f^{}_{\tio}(E^{}_\infty,\omega)|^2$, see Eq.~\eqref{eq:termsinTI}, from terms with $d\equiv|j-l|=0$ through 5 in chains with $L=10$.}
    \label{fig:Sup-OBC-epsilon}
\end{figure}

For the model $\hat H^{}_{\text{OBC}}$ in Eq.~\eqref{eq:H_OBC}, we added a boundary term $\varepsilon \hat S_z^1$ at the first site (with $\varepsilon=0.1$) to break the lattice reflection symmetry. Here, we vary the value of the field strength $\varepsilon$, considering $\varepsilon=0.1,0.01$ and $0.001$ to verify the robustness of the results with respect to the value of $\varepsilon$. 

In Fig.~\ref{fig:Sup-OBC-epsilon}, we compare the results for the spectral functions for the local observable $\hat O^j=\hat S_x^j$ at site $j=\lceil L/2\rceil$ and its translationally invariant counterpart $\hat \tio=\tfrac1L\sum_j \hat S_x^j$ as reported in Fig.~\ref{fig:open-chain} (where $\varepsilon=0.1$), with the results obtained for $\varepsilon=0.01$ and $0.001$, for the largest system size $L=10$ diagonalized. The results for the spectral functions in Fig.~\ref{fig:Sup-OBC-epsilon}(a) and the corresponding site-resolved correlations in Fig.~\ref{fig:Sup-OBC-epsilon}(b) are qualitatively (and quantitatively) similar independently of the values of $\varepsilon$ considered.  

\section{The spectral function $|f^{}_O(E^{}_\infty,\omega)|^2$ versus the ETH function $|\mathrm{f}^{}_O(E^{}_\infty,\omega)|^2$} 

Here we relate the spectral function $|f^{}_O(E^{}_\infty,\omega)|^2$ calculated using the entire energy spectrum and the ETH function $|\mathrm{f}^{}_O(E^{}_\infty,\omega)|^2$ obtained from the average $\overline{|O^{}_{mn}|^2}$ with $\bar E^{}_{mn}\approx E^{}_\infty$ within a microcanonical energy window of width $\delta E$. We consider normalized operators $\hat O$, for example, local operators $\hat O^j$ or normalized translationally invariant operators $\sqrt{L}\hat \tio$. Our goal is to show that, as stated in the main text:
\begin{equation}\label{eq:f_corr and f_var relation}
    |f^{}_O(E^{}_\infty,\omega)|^2\simeq\exp\left(-\frac{\omega^2}{4\sigma_E^2}\right) |\mathrm{f}^{}_O(E^{}_\infty,\omega)|^2\,,\vspace{0.2cm}
\end{equation}
where $\sigma^2_E$ is the energy variance. Therefore, when comparing $|f^{}_O(E^{}_\infty,\omega)|^2$ and $|\mathrm{f}^{}_O(E^{}_\infty,\omega)|^2$ in finite systems, one needs to take into account the relative Gaussian factor $\exp[-\omega^2/(4\sigma^2_E)]$ between the two functions. Since $\sigma^2_E\propto L$, $\exp[-\omega^2/(4\sigma^2_E)]$ approaches unity with increasing $L$, and the functions $|f^{}_O(E^{}_\infty,\omega)|^2$ and $|\mathrm{f}^{}_O(E^{}_\infty,\omega)|^2$ become identical in the thermodynamic limit. Equation~\eqref{eq:f_corr and f_var relation} also holds when $|\mathrm{f}^{}_O(E^{}_\infty,\omega)|^2$ is calculated using the entire spectrum~\cite{patil_rigol_ETH_review_26}.

We first consider $|f^{}_O(E^{}_\infty,\omega)|^2$ coarse-grained over a narrow frequency window of width $\delta \omega$ centered at $\omega$:
\begin{widetext}
\begin{align}\label{eq:f_corr coarsegrained}
    |f^{}_O(E^{}_\infty,\omega)|^2&\simeq\frac{1}{\delta\omega}\int_{\omega-\frac{\delta\omega}{2}}^{\omega+\frac{\delta\omega}{2}} d\omega'|f^{}_O(E^{}_\infty,\omega')|^2 =\frac{1}{D\delta \omega}\sum_{m=1}^D\sum^D_{\substack{n\neq m \\|\omega^{}_{mn}-\omega|\leqslant\frac{\delta\omega}{2}}}|O^{}_{mn}|^2\\&=\frac{1}{D\Omega(E^{}_\infty)}\underbrace{\left[\frac{N^{}_{\delta\omega}(\omega)}{\delta\omega}\right]}_{\rho(\omega)}\left(\frac{\Omega(E^{}_\infty)}{N^{}_{\delta\omega}(\omega)}\sum_{m=1}^{D}\sum^D_{\substack{n\neq m \\ |\omega^{}_{mn}-\omega|\leqslant\frac{\delta\omega}{2}}}|O^{}_{mn}|^2\right)\,\nonumber,
\end{align}
\end{widetext}
\newpage
\onecolumngrid
\noindent where $\Omega(E^{}_\infty)$ is the density of states at the center of the spectrum, $N^{}_{\delta\omega}(\omega)$ is the number of pairs of eigenstates satisfying $|\omega^{}_{mn}-\omega|\leqslant \delta\omega/2$, and $\rho(\omega)$ is the density of states in $\omega$. Using that, in general, the density of states of Hamiltonians with local interactions is Gaussian~\cite{patil_rigol_ETH_review_26}, $\Omega(E)=\tfrac{D}{\sqrt{2\pi \sigma_E^2}}\exp\Big[-\tfrac{(E-E^{}_\infty)^2}{2\sigma_E^2}\Big]\implies \Omega(E^{}_\infty)=\tfrac{D}{\sqrt{2\pi \sigma_E^2}}$,
\begin{equation}\label{eq:densityInOmegafull}
    \rho(\omega)=\int dE \,\Omega(E+\tfrac\omega2)\,\Omega(E-\tfrac\omega2)=\frac{D^2}{2\pi\sigma_E^2} \exp\left(-\frac{\omega^2}{4\sigma_E^2}\right)\int dE \exp\left[-\frac{(E-E^{}_\infty)^2}{\sigma_E^2}\right]=\frac{D^2}{\sqrt{4\pi \sigma_E^2}} \exp\left(-\frac{\omega^2}{4\sigma_E^2}\right).
\end{equation}
Similarly, the density of states in $\omega$ evaluated within a microcanonical energy window of width $\delta E$ centered at $\bar E^{}_{mn}=E^{}_\infty$ is given by
\begin{equation}\label{eq:densityInOmega}
    \rho^{\delta E}(\omega)=\int_{E^{}_\infty-\frac{\delta E}{2}}^{E^{}_\infty+\frac{\delta E}{2}} dE \,\Omega(E+\tfrac\omega2)\,\Omega(E-\tfrac\omega2)=\rho(\omega)\left(\frac{1}{\sqrt{\pi \sigma_E^2}}\int_{-\frac{\delta E}{2}}^{\frac{\delta E}{2}} dE' \exp\left[-\frac{E'^2}{\sigma_E^2}\right]\right)=\rho(\omega)\,\mathrm{erf}\left(\frac{\delta E}{2\sigma^{}_E}\right)\,,
\end{equation}
where $\mathrm{erf}(z)=\int_{-z}^{z} dt \exp(-t^2)/\sqrt{\pi}$ is the error function.

Next, we compute the sum over the matrix elements in Eq.~\eqref{eq:f_corr coarsegrained}, restricting it to a microcanonical energy window characterized by the indicator function $\mathbf{1}^{}_{|\bar E^{}_{mn}-E^{}_\infty|\leqslant \delta E/2}\equiv \mathbf{1}^{}_{\delta E}$. Using the ETH ansatz~\eqref{eq:ETH}, we find:
\begin{align}\label{eq:SumMatMicro}
    &\sum_{m=1}^{D}\sum^D_{\substack{n\neq m \\ |\omega^{}_{mn}-\omega|\leqslant\frac{\delta\omega}{2}}}|O^{}_{mn}|^2\,\mathbf{1}^{}_{\delta E}=\delta\omega\int_{E^{}_\infty-\frac{\delta E}{2}}^{E^{}_\infty+\frac{\delta E}{2}} dE \,\frac{\Omega(E+\tfrac\omega2)\,\Omega(E-\tfrac\omega2)}{\Omega(E)}|\mathrm{f}^{}_O(E,\omega)|^2\\
    &=D\delta\omega\exp\left(-\frac{\omega^2}{4\sigma_E^2}\right)\int_{-\frac{\delta E}{2}}^{\frac{\delta E}{2}}dE' \frac{e^{-E'^2/(2\sigma_E^2)}}{\sqrt{2\pi\sigma_E^2}}|\mathrm{f}^{}_O(E^{}_\infty+E',\omega)|^2\simeq D\delta\omega\exp\left(-\frac{\omega^2}{4\sigma_E^2}\right)|\mathrm{f}^{}_O(E^{}_\infty,\omega)|^2\,\mathrm{erf}\left(\frac{\delta E}{2\sqrt{2}\sigma^{}_E}\right)\,,\nonumber
\end{align}
where, in the last step, we retained the leading contribution to the integral by Taylor expanding $\mathrm{f}^{}_O(E,\omega)$ about $E=E^{}_\infty$, assuming $\delta E\ll \sigma^{}_E$ and using the subextensive scaling $\sigma^{}_E\propto \sqrt{L}$~\cite{patil_rigol_ETH_review_26}. Equation~\eqref{eq:SumMatMicro} shows that the sum within a microcanonical window is related to the sum over the entire spectrum in Eq.~\eqref{eq:f_corr coarsegrained} via
\begin{equation}\label{eq:ResultSumMatMicro}
    \sum_{m=1}^{D}\sum^D_{\substack{n\neq m \\|\omega^{}_{mn}-\omega|\leqslant\frac{\delta\omega}{2}}}|O^{}_{mn}|^2 \,\mathbf{1}^{}_{\delta E}\simeq \left[\sum_{m=1}^{D}\sum^D_{\substack{n\neq m \\|\omega^{}_{mn}-\omega|\leqslant\frac{\delta\omega}{2}}}|O^{}_{mn}|^2\right] \mathrm{erf}\left(\frac{\delta E}{2\sqrt{2}\sigma^{}_E}\right)\,.
\end{equation}

Hence, the rescaled variance in Eq.~\eqref{eq:f_corr coarsegrained} calculated over the entire spectrum can be related to that calculated within a microcanonical window, \ie the ETH spectral function $|\mathrm{f}^{}_O(E^{}_\infty,\omega)|^2$, using the ratio of Eqs.~\eqref{eq:ResultSumMatMicro} and \eqref{eq:densityInOmega}:
\begin{align}\label{eq:rescaled variance}
    &\frac{\Omega(E^{}_\infty)}{\rho(\omega)\delta\omega}\sum_{m=1}^{D}\sum^D_{\substack{n\neq m \\|\omega^{}_{mn}-\omega|\leqslant\frac{\delta\omega}{2}}}|O^{}_{mn}|^2\simeq \frac{\mathrm{erf}\left(\frac{\delta E}{2\sigma^{}_E}\right)}{\mathrm{erf}\left(\frac{\delta E}{2\sqrt{2}\sigma^{}_E}\right)} \left(\frac{\Omega(E^{}_\infty)}{\rho^{\delta E}(\omega)\delta\omega}\sum_{m=1}^{D}\sum^D_{\substack{n\neq m \\|\omega^{}_{mn}-\omega|\leqslant\frac{\delta\omega}{2}}}|O^{}_{mn}|^2 \,\mathbf{1}^{}_{\delta E}\right)\nonumber\\
    &= \frac{\mathrm{erf}\left(\frac{\delta E}{2\sigma^{}_E}\right)}{\mathrm{erf}\left(\frac{\delta E}{2\sqrt{2}\sigma^{}_E}\right)} |\mathrm{f}^{}_O(E^{}_\infty,\omega)|^2\simeq \sqrt{2}\, |\mathrm{f}^{}_O(E^{}_\infty,\omega)|^2\,,
\end{align}
where, in the last step, we used that $\mathrm{erf}(z)\simeq \tfrac{2z}{\sqrt{\pi}}$ for $z\ll 1$. Substituting $\Omega(E^{}_\infty)$, $\rho(\omega)$ from~\eqref{eq:densityInOmegafull}, and the rescaled variance from Eq.~\eqref{eq:rescaled variance} into Eq.~\eqref{eq:f_corr coarsegrained}, one obtains the relation in Eq.~\eqref{eq:f_corr and f_var relation}.

%%%%%%%%%%%%%%%%%%    End Supplementary    %%%%%%%%%%%%%%%%%%%%

\end{document}